\documentclass[11pt,preprint]{aastex}

\begin{document}

\title{Magnetic Field Amplification by Turbulence in A Relativistic Shock Propagating through An Inhomogeneous Medium}

\author{
Yosuke Mizuno\altaffilmark{1,2}, Martin Pohl\altaffilmark{3,4}, Jacek Niemiec\altaffilmark{5}, Bing Zhang\altaffilmark{6}, Ken-Ichi Nishikawa\altaffilmark{1,2}, and Philip E. Hardee\altaffilmark{7} }

\altaffiltext{1}{Center for Space
Plasma and Aeronomic Research, University of Alabama in Huntsville, 
320 Sparkman Drive, NSSTC, Huntsville, AL 35805, USA; mizuno@cspar.uah.edu}
\altaffiltext{2}{National Space Science and Technology Center,
VP62, Huntsville, AL 35805, USA}
\altaffiltext{3}{Institut fur Physik und Astronomie, Universit\"{a}t Potsdam, 14476 Potsdam-Golm, Germany}
\altaffiltext{4}{DESY, Platanenallee 6, 15738 Zeuthen, Germany}
\altaffiltext{5}{Institute of Nuclear Physics PAN, ul. Radzikowskiego 152, 31-342 Krak\'{o}w, Poland}
\altaffiltext{6}{Department of Physics and Astronomy, University of Nevada, Las Vegas, NV 89154, USA}
\altaffiltext{7}{Department of
Physics and Astronomy, The University of Alabama, Tuscaloosa, AL
35487, USA}

\shorttitle{Turbulent Magnetic field amplification in Relativistic Shocks}
\shortauthors{Mizuno et al.}

\begin{abstract}

We perform two-dimensional relativistic magnetohydrodynamic simulations of a mildly relativistic shock propagating through an inhomogeneous medium. We show that the postshock region becomes turbulent owing to preshock density inhomogeneity, and the magnetic field is strongly amplified due to the stretching and folding of field lines in the turbulent velocity field. The amplified magnetic field evolves into a filamentary structure in two-dimensional simulations. The magnetic energy spectrum is flatter than the Kolmogorov spectrum and indicates that a so-called small-scale dynamo is occurring in the postshock region. We also find that the amount of magnetic-field amplification depends on the direction of the mean preshock magnetic field, and the time scale of magnetic-field growth depends on the shock strength.

\end{abstract}

\section{Introduction}

Relativistic shocks accelerate plasma to ultrarelativistic velocities. Emission from such shocks is observed in astrophysical sources on many length, time and energy scales, and is typically attributed to the synchrotron process. The composition of the plasma, the magnitude of the preshock magnetic field, and the Lorentz factor of the shock are generally unknown. One of the best-studied systems is the radiative afterglow that follows the bright flash of $\gamma$-rays in gamma-ray bursts (GRBs). The structure of the emitting region can be modeled from the observed emission: in the standard GRB afterglow model (e.g., Piran 2005; M\'{e}sz\'{a}ros 2006), the radiation is produced in a relativistic blastwave shell propagating into a weakly magnetized plasma. Detailed studies of GRB spectra and light curves have shown that the magnetic energy density in the emitting region is a fraction $\epsilon_{B} \sim 10^{-3} - 10^{-1}$ of the internal energy density (e.g., Panaitescu \& Kumar 2002, Yost et al. 2003; Panaitescu 2005). However, simple compressional amplification of the weak pre-existing magnetic field of the circumburst medium can not account for such high magnetization (Gruzinov 2001).

The leading hypothesis for field amplification in GRB afterglows is the relativistic Weibel instability that produces filamentary currents aligned with the shock normal; these currents are responsible for the creation of transverse magnetic fields (e.g., Gruzinov \& Waxman 1999; Medvedev \& Loeb 1999). Recent plasma simulations using the particle-in-cell (PIC) method have demonstrated magnetic-field generation in relativistic collisionless shocks (e.g., Nishikawa et al. 2003, 2009; Spitkovsky 2008). However, the size of the simulated regions is orders of magnitude smaller than the GRB emission region. It remains unclear whether magnetic fields generated on scales of tens of plasma skin depths will persist at sufficient strength in the entire emission region, some $10^6$ plasma skin depths in size. On the other hand, in magnetohydrodynamic (MHD) processes, if the density of the preshock medium is strongly inhomogeneous, significant vorticity is produced in the shock transition that stretches and deforms magnetic field lines leading to field amplification (Goodman \& MacFadyen 2007; Sironi \& Goodman 2007; Palma et al.\ 2008). For the long duration GRBs, i.e., associated with the iron core collapse of mass-losing very massive stars, density fluctuations in the interstellar medium may arise through several processes (e.g., Sironi \& Goodman 2007).

There is also a direct observational motivation for relativistic turbulence in GRB outflows. Significant angular fluctuations by relativistic turbulence have been invoked to explain the large variation of the $\gamma$-ray luminosity among bursts as well as the intraburst variability observed in many afterglows. The afterglow polarization indicates a breaking of axial symmetry; the correlation of this polarization with large-amplitude variability suggests blastwave anisotropy as the source of variability. Narayan \& Kumar (2009) and Lazar et al. (2009) have proposed a relativistic turbulence model instead of the well-known internal shock model as the production mechanism for variable GRB light curves and applied it to GRB 080319B (Kumar \& Narayan 2009). Zhang \& Yan (2010) have developed a new GRB prompt-emission model in the highly magnetized regime, namely, an internal-collision-induced magnetic reconnection and turbulence (ICMART) model. 
Within this model the short-time variability ``spikes'' are related to turbulence in the magnetic dissipation region while the broad variability component in the GRB lightcurves is related to central engine activity.

The innermost parts of the relativistic jets seen in Active Galactic Nuclei (AGN) are explored by multi-wavelength observational campaigns that provide probes into the structure of the blazar zone (e.g., Marscher et al.\ 2008, 2010). Recently correlated X-ray/TeV gamma-ray flares with timescales from 15 minutes (Mrk 421) to a few hours (Mrk 501 and 1ES 1959+650) have been observed (Aharonian et al.\ 2003; Albert et al.\ 2007; Krawczynski et al.\ 2004). The fast variable flares may come from small regions, a few Schwarzschild radii in size. Several authors have proposed fast-moving needles within a slower jet or a jet-within-a-jet scenario to explain the fast variability of blazars (Levinson 2007; Begelman et al.\ 2008; Ghisellini \& Tavecchio 2008; Giannios et al.\ 2008). On the other hand, Marscher \& Jorstad (2010) have proposed that some short-term fluctuations can be understood as the consequence of a turbulent ambient-jet-plasma that passes through shocks in the jet flow (Marscher et al.\ 1992).

Supernova remnants (SNRs) are bound by an expanding nonrelativistic spherical blastwave. The synchrotron emission from SNRs is generally consistent with compression of the interstellar field. However, recent discovery of the year-scale variability in the synchrotron X-ray emission of SNRs suggests that the magnetic field needs to be amplified in the SNR up to the milli-Gauss level (Uchiyama et al.\ 2007, Uchiyama \& Aharonian 2008), although the interpretation is disputed (e.g., Bykov et al.\ 2008, Huang \& Pohl 2008). The evidence for magnetic field amplification in SNRs has been obtained also from the thickness of X-ray filaments that typically indicate $\sim 100$ micro-Gauss magnetic fields (Bamba et al.\ 2003, 2005a, 2005b; Vink \& Laming 2003, but see also Pohl et al.\ 2005). Since the typical magnetic-field strength in the interstellar medium is on the order of a few micro-Gauss, amplification beyond simple shock compression is necessary to achieve a milli-Gauss level magnetic field in SNRs.

Recently, Giacalone \& Jokipii (2007) have performed non-relativistic MHD shock simulations including density fluctuations with a Kolmogorov power spectrum in the preshock medium. They observed a strong magnetic-field amplification caused by turbulence in the postshock medium; the final rms magnetic-field strength is reportedly a hundred times larger than the preshock field strength. 
Density fluctuations with a Kolmogorov power spectrum would be present in the preshock medium from the transonic turbulence expected in the diffuse interstellar medium (ISM) (e.g., Hennebelle \& Audit 2007, Inoue \& Inutsuka 2008, Inoue et al.\ 2009). Inoue et al.\ (2009) have also obtained strong magnetic-field amplification by turbulence associated with a strong thermal instability driven shock wave propagating through an inhomogeneous interstellar two-phase medium. 
The interaction of SNRs with a turbulent, magnetized ISM has been studied by Balsara et al.\ (2001) via three-dimensional MHD simulations. They found that the interaction of strong shocks with the turbulent density fluctuations could be a strong source of helicity generation which provides a potential source of magnetic field amplification. Kim et al.\ (2001) studied the energetics, structure, and spectra of a supernova (SN) driven turbulent medium with higher resolution simulations. Balsara et al.\ (2004) have investigated in detail the role of magnetic field amplification in SN-driven turbulence and found that the helicity generation mechanism by SN-driven, turbulent, multiphase ISM processes was adequate to sustain the magnetic field growth (also Balsara \& Kim 2005; Kim \& Balsara 2006). Haugen et al.\ (2004) have performed simulations and found that magnetic field amplification can be sustained by shock-driven turbulence in an isothermal, compressible, shock-driven medium. All the above simulations are relevant to SNR shocks, which are non-relativistic. GRBs and AGN jets have relativistic shocks. Similar physical processes should apply, so that these shocks would also experience strong magnetic-field amplification by turbulence.

The magnetic-field amplification mechanism is by a so-called turbulent dynamo or small-scale dynamo, also known as a fast dynamo, and is different from the classical mean-field dynamo mechanism. Mean-field dynamo theory (e.g., Steenbeck et al.\ 1966; Parker 1971; Ruzmaikin et al.\ 1988, Diamond et al.\ 2005) has been used to study the growth of the large-scale magnetic field in our Galaxy. In general, mean-field dynamos are slow, involve almost incompressible motions, and the growth rate decreases with decreasing resistivity ($\nu$). Kulsrud \& Anderson (1992) pointed out that the magnetic Reynolds number ($\mbox{Re}_{m} \equiv \ell v/\nu$, where $\ell$ is typical length of flow) in our Galaxy is very large ($\mbox{Re}_{m} \sim 10^{16}$), and such a high $\mbox{Re}_{m}$ leads to the development of turbulent velocity and magnetic-field fluctuations on small scales. While the growth of large-scale magnetic fields is probably driven by a mechanism similar to a mean-field dynamo, other mechanisms are likely responsible for the more rapid generation of field on smaller scales. The theory of fast dynamos  has been developed to investigate the rapid growth of magnetic field on smaller scales in the limit of infinite magnetic Reynolds number, and the growth rate does not depend on resistivity (e.g., Childress \& Gilbert 1995; Galloway 2003; Brandenburg \& Subamanian 2005). Here the amplification of the magnetic field arises from sequences of a ``stretch, twist, and fold'' nature. While the limit ($\mbox{Re}_{m}=\infty$) is unrealistic in astrophysical simulations, published results of incompressible MHD simulations (e.g., Schekochihin et al.\ 2004) and compressible MHD simulations (e.g., Balsara et al.\ 2004; Balsara \& Kim 2005; Kim \& Balsara 2006) are nevertheless in agreement with the existence of a small-scale turbulent driven dynamo.

Here we report on magnetic-field amplification by turbulence in two-dimensional relativistic magnetohydrodynamic (RMHD) simulations of a mildly relativistic shock wave propagating through an inhomogeneous medium. In this paper we focus on the small-scale growth of the magnetic field.
This paper is organized as follows: We describe the numerical method and setup used for our simulations in \S 2, present our results in \S 3, and discuss the astrophysical implications in \S 4.

\section{Numerical Method and Setup}

To study the time evolution of a mildly relativistic shock propagating in an inhomogeneous medium, we use the 3D GRMHD code ``RAISHIN'' in two-dimensional Cartesian geometry ($x-y$ plane). RAISHIN is based on a $3+1$ formalism of the general relativistic conservation laws of particle number and energy-momentum, Maxwell's equations, and Ohm's law with no electrical resistance (ideal MHD condition) in a curved spacetime (Mizuno et al.\ 2006)
\footnote{
Constained transport schemes are used to maintain divergence-free magnetic field in the RAISHIN code. This scheme requires the magnetic field to be defined at the cell interfaces. On the other hand, conservative, high-resolution shock capturing schemes (Godonov-type schemes) for convservation laws require the variables to be defined at the cell center. In order to combine variables defined at these different positions, the magnetic field calculated at the cell interfaces is interpolated to the cell center and as a result the scheme becomes
non-conservative even though we solve the conservation laws (Komissarov 1999).}.
In the RAISHIN code, a conservative, high-resolution shock-capturing (HRSC) scheme is employed. The numerical fluxes are calculated using the HLL approximate Riemann solver, and flux-interpolated constrained transport (flux-CT) is used to maintain a divergence-free magnetic field. The RAISHIN code performs special relativistic calculations in Minkowski spacetime by choosing the appropriate metric. RAISHIN has proven to be accurate to second order and has passed a number of one-dimensional and multidimensional numerical tests including highly relativistic cases and highly magnetized cases in both special and general relativity (Mizuno et al.\ 2006; Mizuno et al. 2010).  We have improved the reconstruction schemes in the RAISHIN code to handle the turbulent structure more finely. A fifth-order weighted essentially non-oscillatory (WENO, Jiang \& Shu 1996) scheme is built into the code and used in the simulations. WENO schemes provide highly-accurate solutions in regions of smooth flow and non-oscillatory transitions in the presence of discontinuous waves by combining different interpolation stencils of order $r$ into a weighted average of order $2r-1$.

At the beginning of the simulations, an inhomogeneous plasma with mean rest-mass density $\rho_{0}$ and containing  fluctuations $\delta \rho$ is established across the whole simulation region and uniformly flows in the positive $x$-direction with speed $v_{0}$, where $\rho_{0}$ is an arbitrary normalization constant (our simulations are scale free) and we choose $\rho_{0}=1.0$.
The density fluctuations are generated according to a two-dimensional Kolmogorov-like power-law spectrum of the form 
\begin{equation}
P_{k} \propto {1 \over 1 + (kL)^{8/3}},
\end{equation}
where $k$ is the wave number and $L$ is the turbulence coherence length.

A method used to generate fluctuations with a Kolmogorov-like turbulence spectrum was proposed in Giacalone \& Jokipii (2007) (also Giacalone \& Jokipii 1999). Following their method, our turbulence is described by summing over a large number of discrete wave modes,
\begin{equation}
\delta \rho (x,y)=\sum^{N_{m}}_{n=1}A(k_{n})\,\exp[i(k_{n}\,\cos \theta_{n} x +
k_{n} \,\sin \theta_{n} y + \phi_{n})],
\end{equation}
where $A(k_{n})$ is the amplitude, $\theta_{n}$ is the direction of the wavevector with magnitude $k_{n}$, and $\phi_{n}$ is the phase. For each $n$ a phase and a direction are randomly selected from the ranges $0 < \phi_{n} < 2 \pi$ and $0 < \theta_{n}< 2 \pi$, respectively. The total number of modes is $N_{m}=50$. The amplitude in the equation above is given by
\begin{equation}
A^{2}(k_{n})=a_{0} G(k_{n}) \left [ \sum^{N_{m}}_{n=1} G(k_{n}) \right]^{-1},
\end{equation}
and
\begin{equation}
G(k_{n}) = {2 \pi k_{n} \Delta k_{n} P(k)},
\end{equation}
where $a_{0}$ is the wave variance. In the simulations we choose $a_{0}=0.02$, which makes a fluctuation variance of $\sqrt{<\delta \rho^{2}>} = 0.012 \rho_{0}$.
We choose the maximum wavelength as $\lambda_{max}=0.5L$, one-half of the size of the simulation box in the $y$-direction, and the minimum wavelength as $\lambda_{min}=0.025L$. 

We consider a low gas-pressure medium with constant $p=0.01\rho_{0}c^{2}$, where $c$ is the speed of light. The equation of state is that of an ideal gas with $p=(\Gamma-1) \rho e$, where $e$ is the specific internal energy density and the adiabatic index $\Gamma=5/3$. The specific enthalpy is $h \equiv 1+e/c^2 + p/ \rho c^{2}$.

The pre-shock plasma carries a weak constant magnetic field. The magnetic field amplitude is $B_{0}=4.5 \times 10^{-3} \sqrt{4 \pi \rho_{0} c^{2}}$ which leads to a high plasma-$\beta$ ($\beta = p_{gas}/p_{mag}=10^{3}$). To investigate the effect of the magnetic field direction with respect to the shock propagation direction, we choose two different directions, magnetic field parallel ($B^{x}$, $\theta_{Bn}=0\arcdeg$) and perpendicular ($B^{y}$, $\theta_{Bn}=90\arcdeg$) to the shock normal. Using the mean plasma density ($\rho_{0}$), the sound speed $c_{s}/c=(\Gamma p/\rho h)^{1/2}$ and the Alfv\'{e}n speed $v_{A}/c=[B^2/(\rho h + B^2)]^{1/2}$ are $c_s = 0.128c$ and $v_A = 0.0047c$.

\begin{deluxetable}{lcccc}
\tablecolumns{7}
\tablewidth{0pc}
\tablecaption{Models and Parameters}
\label{table1}
\tablehead{
\colhead{Case} & \colhead{$v_{0}$} & \colhead{$\theta_{Bn}$} &
\colhead{$t_{max}/c$} & \colhead{$N/L$}
}
\startdata
A & 0.4 & 0 & 10 & 256 \\
B & 0.4 & 90 & 10  & 256 \\
C & 0.9 & 0 & 5   & 256 \\
D & 0.9 & 90 & 5   & 256 \\
\enddata
\end{deluxetable}
The size of the computational domain is $(x,y)=(2L, L)$ with an $N/L=256$ grid resolution. We impose periodic boundary conditions in the $y$-direction. In order to create a shock wave, a rigid reflecting boundary is placed at $x=x_{max}$. The fluid, which is moving initially in the positive $x$-direction, is stopped at this boundary, where the velocity $v^x$ is set to zero. As the density builds up at $x=x_{max}$, a shock forms and propagates in the $-x$-direction. In the frame of reference of the simulation (contact-discontinuity frame), the plasma velocity downstream of the shock is zero on average.  New fluid continuously flows in from the inner boundary ($x=0$) and density fluctuations are advected with the flow speed. In the frame of reference of the inflowing fluid the density fluctuation pattern is maintained at each time step using eq.(2) but extended in the -x direction at the flow speed, $v_0$, so that density fluctuations are inserted in the inflowing material at the inner simulation boundary. We choose two different flow speeds, $v^{x}=v_{0}=0.4c$ and $0.9c$. The various different cases that we have considered are listed in Table 1.

\section{Results}

\subsection{Shock structure}

Figures 1 and 2 show 2-D images of the density and the total magnetic field strength at $t_{s}=5.0$ and $t_{s}=10.0$, where $t_{s}$ is in units of $L/c$ with $c=1$, for the parallel ($B^{x}$) and perpendicular ($B^{y}$) magnetic field cases with slow flow velocity $v_{0}=0.4c$
\begin{figure}[h!]
\epsscale{0.9}
\plotone{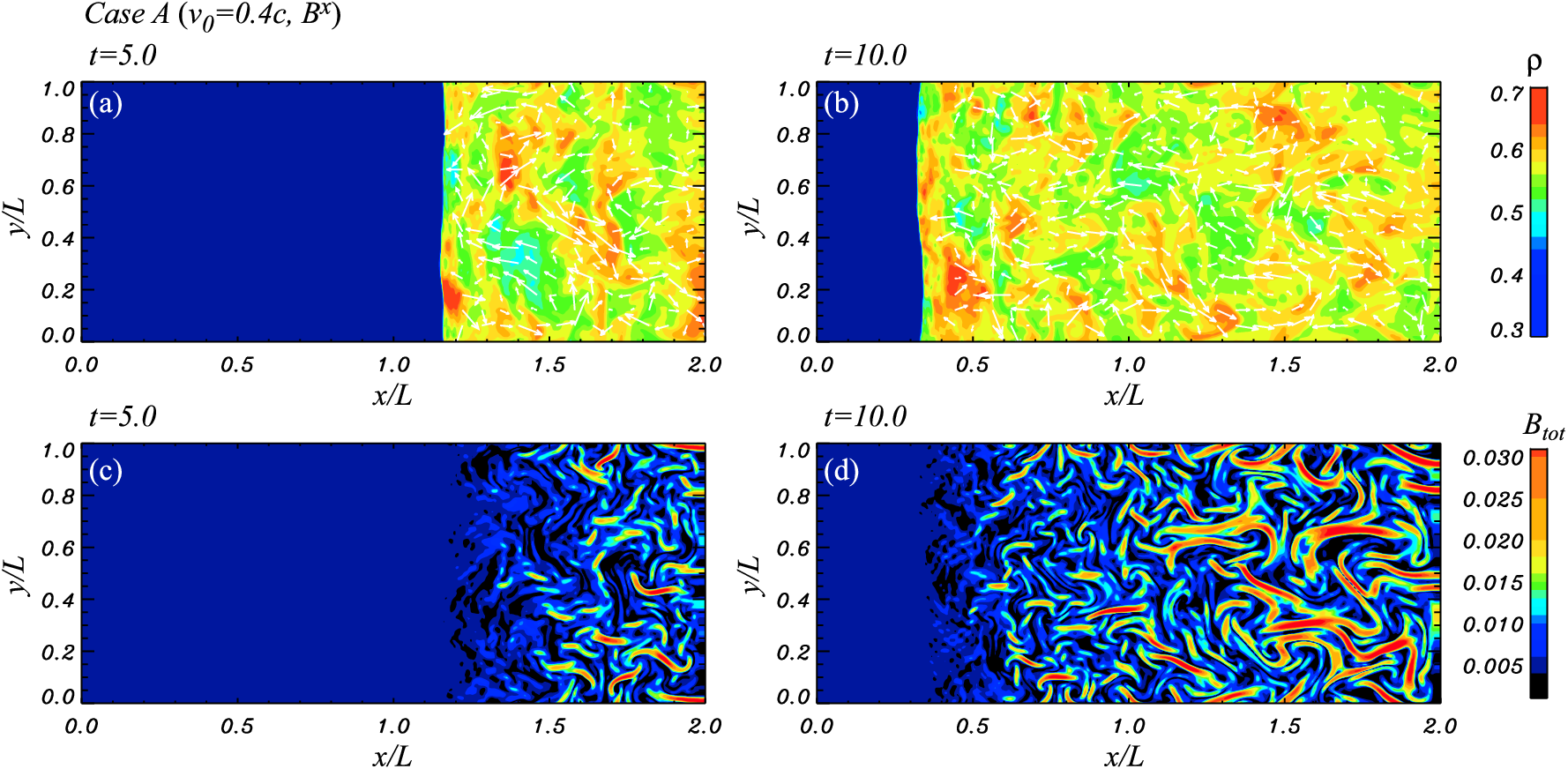}
\caption{Two-dimensional images of density ({\it upper}) and total magnetic field strength ({\it lower}) at $t_{s}=5.0$({\it left}) and $t_{s}=10.0$({\it right}) for case A (magnetic field parallel to the shock propagation direction and $v_{0}=0.4c$). White arrows indicate the flow direction in the postshock region.
\label{f1}}
\end{figure}
When the preshock plasma with inhomogeneous density encounters the shock, the shock front is rippled, leading to significant, random transverse flow behind the shock. 
These ripples in the shock front introduce rotation and vorticity in the postshock region through a process similar to the Richtmyer-Meshkov instability (e.g., Brouillette 2002). This instability is excited when a sudden density jump encounters a shock and is a possible source of vorticity generation (e.g., Zabusky 1999)
\footnote{
Samtaney \& Zabusky (1994) calculated scaling relations for vorticity generation via the Richtmyer-Meshkov instability when shocks interact with density inhomogeneities in two dimensions and found that increasing shock Mach number or density contrast enhances the generation of vorticity at the shocks. }.
Our simulation starts from preexisting fluctuations having a broad range of scales in the preshock medium. The flow pattern in the postshock region is initially highly nonlinear and comparison with linear instability analysis is not useful.
\begin{figure}[h!]
\epsscale{0.9}
\plotone{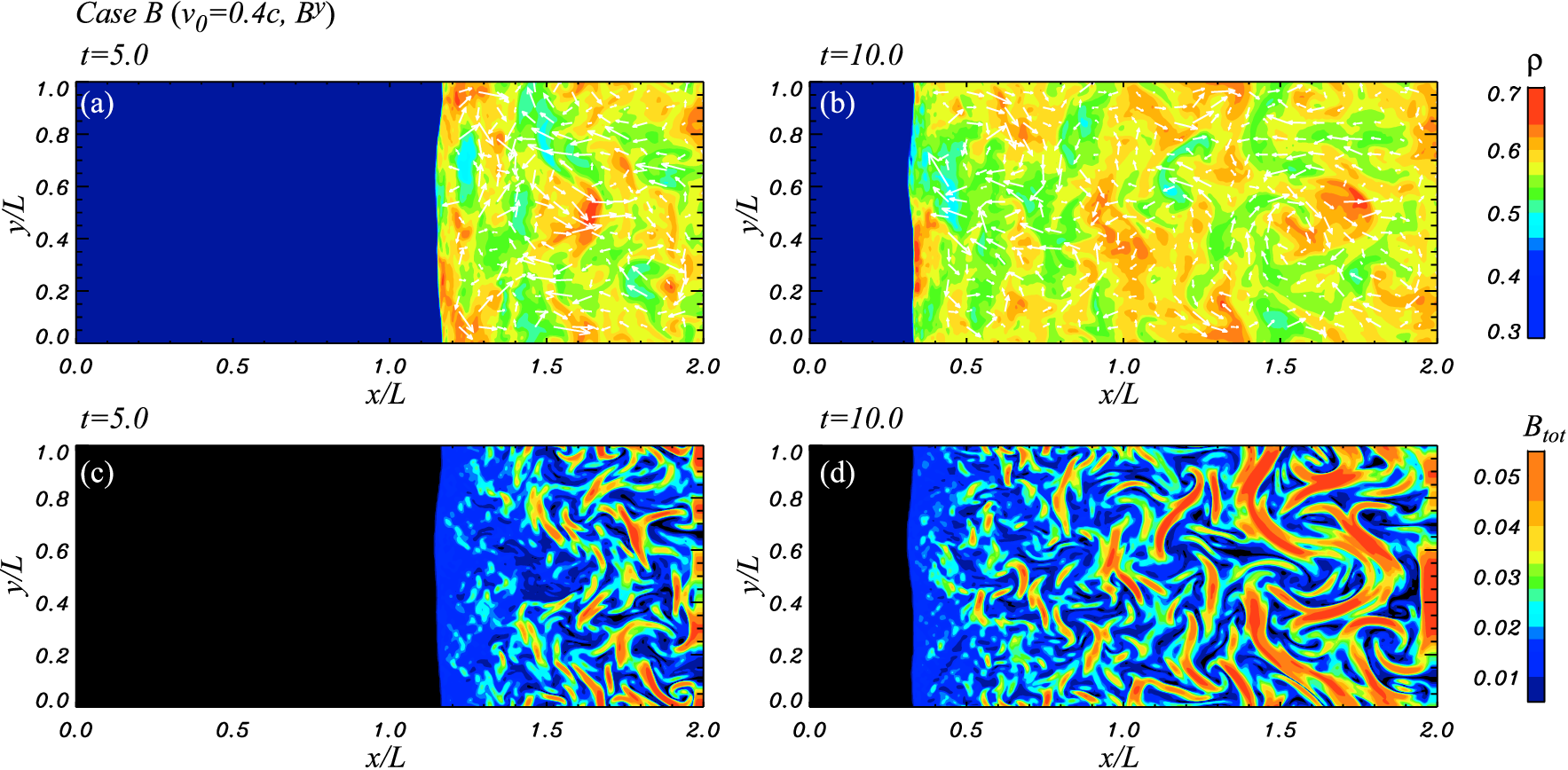}
\caption{Two-dimensional images of density ({\it upper}) and total magnetic field strength ({\it lower}) at $t_{s}=5.0$({\it left}) and $t_{s}=10.0$({\it right}) for case B (magnetic field perpendicular to the shock propagation direction and $v_{0}=0.4c$). White arrows indicate the flow direction in the postshock region.
\label{f2}}
\end{figure}

Since the preexisting magnetic field is much weaker than the postshock turbulence, the turbulent velocity field can easily stretch and deform the frozen-in magnetic field, resulting in its distortion and amplification. This creates regions with higher magnetic field intensity. Near the shock front, the vorticity scale size is small, but far away from the shock front the vorticity scale size becomes larger through an inverse cascade of vortices, and the magnetic field is strongly amplified with time. The amplified magnetic field evolves into a filamentary structure. 
We note that the magnetic-field amplification under study here does not increase the total magnetic flux, only the magnetic-field strength.
The average turbulent velocity in the postshock region is $ \sim 0.02c$ at $t_{s}=10$ in both cases A and B. The average sound speed and Alfv\'{e}n velocity in the postshock region at $t_{s}=10$ are $\sim 0.32c$ and $\sim 0.012c$, respectively in case A and $\sim 0.33c$ and $\sim 0.0044c$, respectively in case B. Thus the turbulent velocity is subsonic and super-Alfv\'{e}nic in most of the postshock region in both cases A and B. 
This result is consistent with previous non-relativistic studies (e.g., Balsara et al.\ 2001; Kim et al.\ 2001; Balsara et al.\ 2004; Kim \& Balsara 2006; Haugan et al.\ 2004, Giacalone \& Jokipii 2007; Inoue et al.\ 2009), although the turbulent velocity in the postshock region is locally supersonic in these previous non-relativistic simulations.

Figure 3 shows one-dimensional profiles along the $x$-axis at $y/L=0.5$ at $t_{s}=10$ for cases A and B, involving slow flow velocity and parallel or perpendicular magnetic field. 
\begin{figure}[h!]
\epsscale{0.85}
\plotone{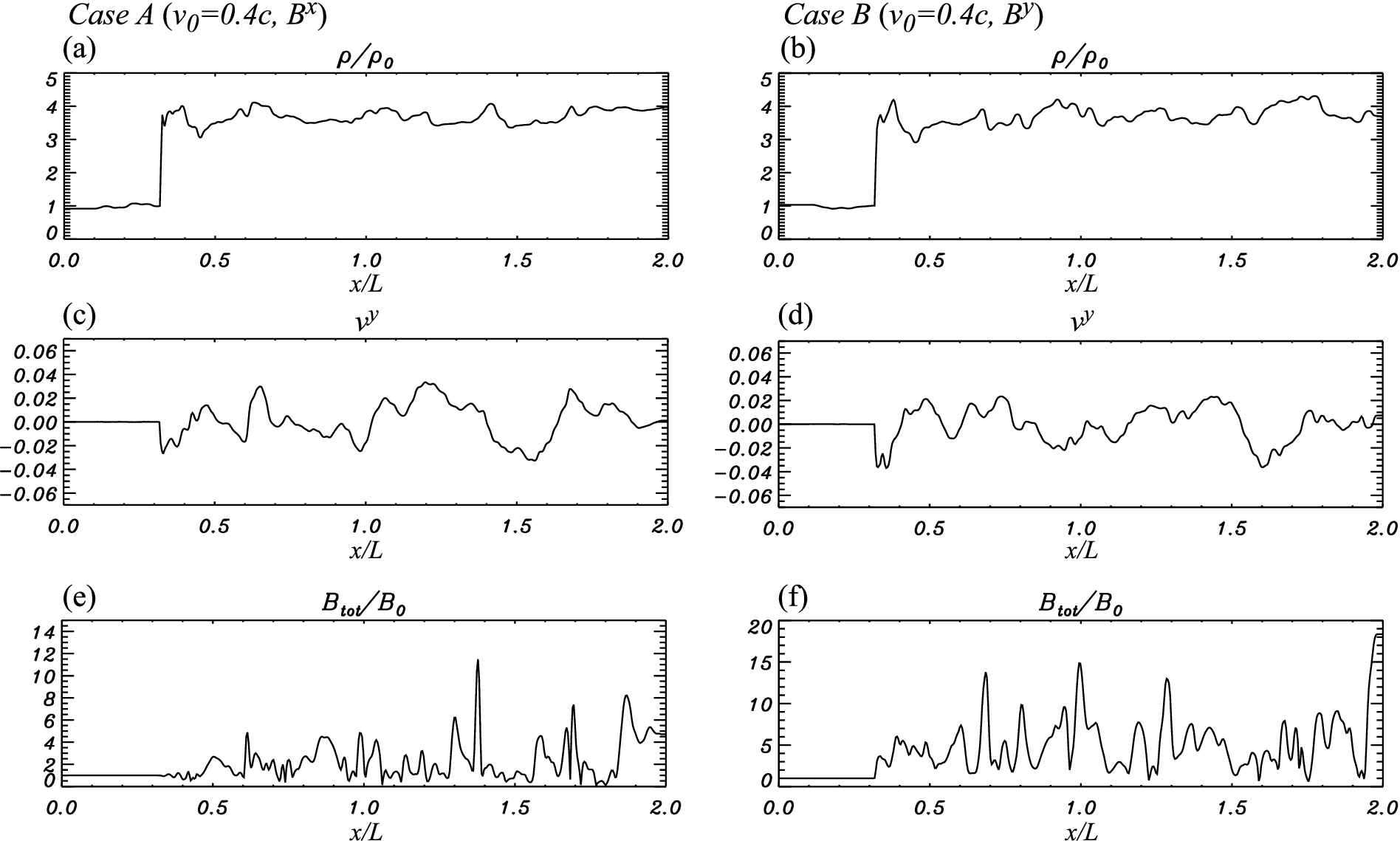}
\caption{One-dimensional profile of ({\it panels a, b}) the density normalized by the initial mean density ($\rho_{0}$), ({\it panels c, d}) the transverse velocity ($v^{y}$), and ({\it panels e, f}) the total magnetic field strength ($B_{tot}$) normalized by the initial magnetic field ($B_{0}$), all shown along the $x$-direction at $y/L=0.5$ and $t_{s}=10.0$ for case A ({\it left}, a parallel initial magnetic field and $v_{0}=0.4c$), and case B ({\it right}, a perpendicular initial magnetic field and $v_{0}=0.4c$). The shock front is located at $x/L=0.31$. The left and the right sides of shock front are upstream and downstream regions, respectively.
\label{f3}}
\end{figure}
The density, normalized to $\rho_0$, the transverse velocity ($v^{y}$), and the total magnetic field strength are plotted from top to bottom. The shock front is located at $x=0.31L$. The left and right sides of the shock front are upstream and downstream, respectively. The shock propagation speed is around $v_{sh} \sim 0.17c$, in good agreement with analytical calculations (see the Appendix).  We parameterize the shock strength by the relativistic Mach number $M_{s} \equiv \gamma'_{sh} v'_{sh}/\gamma_{s} c_{s}$, where $\gamma'_{sh}$ and $v'_{sh}$ are the shock propagation speed and the shock Lorentz factor measured in the inflow frame (see the Appendix) and $\gamma_{s} \equiv (1-c_{s}^{2})^{-1/2}$ is the Lorentz factor associated with the sound speed. The shock propagation speed obtained from the simulations and the sound speed in the preshock region leads to $M_{s} \sim 4.9$ in the slow flow case.  The density jumps by nearly a factor of 4, which is the strong-shock limit in the Newtonian case. The transverse-velocity profiles show strongly fluctuating structure. The maximum transverse velocity is about $0.04c$ in both cases, which is subsonic in the postshock region ($<c_{s}> \simeq 0.32c$ in slow flow cases). The total-magnetic-field profiles also show strong fluctuations. In the perpendicular field case, the magnetic field is compressed by nearly a factor 3.5 behind the shock front, but further downstream the magnetic field locally reaches more than 10 times the amplitude of the initial field. 
As a result of the density fluctuations the magnetic field compression behind the shock front is not identical to the density jump.  In the parallel-field case, magnetic compression at the shock does not occur. Instead, the magnetic field is amplified purely by turbulent motion, locally reaching more than 5 times the initial field amplitude.

\begin{figure}[h!]
\epsscale{0.9}
\plotone{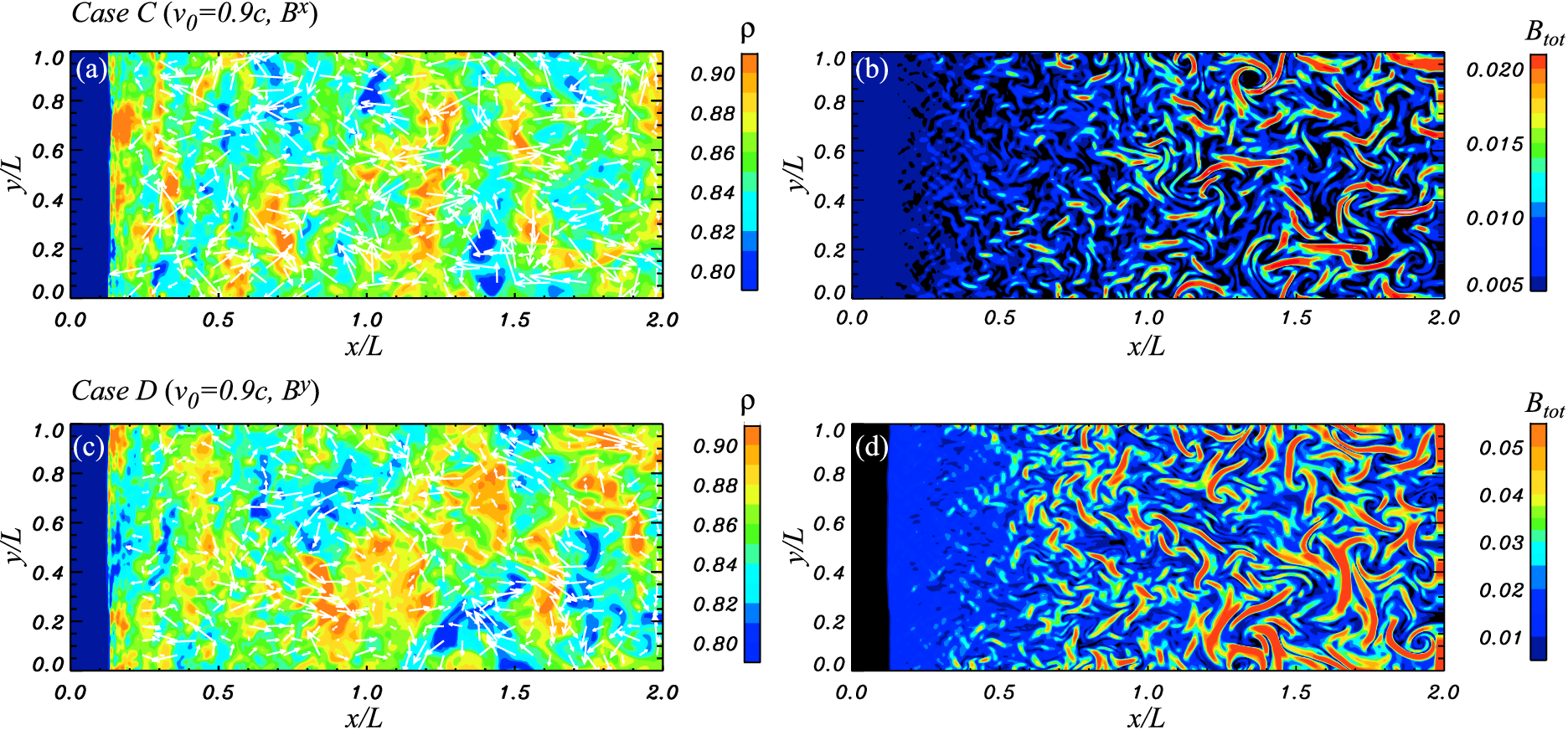}
\caption{Two-dimensional images of the density ({\it left})
and the total magnetic-field strength ({\it right}) at $t_{s}=4.4$ for the case C ({\it upper}, parallel magnetic field, $v_{0}=0.9c$) and D ({\it lower}, perpendicular magnetic field, $v_{0}=0.9c$). White arrows indicate the flow direction in the postshock region.
\label{f4}}
\end{figure}
Figure 4 shows a 2-D image of the density and the total magnetic-field strength at $t_{s}=4.4$ for the parallel ($B^{x}$) and perpendicular ($B^{y}$) magnetic field cases with fast flow velocity, $v_{0}=0.9c$. When the flow speed increases, the shock becomes stronger. The shock propagation speed for $v_{0}=0.9c$ is more than two times faster than that in $v_{0}=0.4c$ cases. In the postshock region, turbulent motion is created by ripples in the shock front, and the magnetic field is deformed and amplified by the turbulence. Filamentary magnetic fields are seen also in these cases. 
The average turbulent velocity in the postshock region is $ \sim 0.035c$ at $t_{s}=4.4$ in both cases C and D. The average sound speed and Alfv\'{e}n velocity in the postshock region at $t_{s}=4.4$ are $\sim 0.68c$ and $\sim 0.0063c$ in case C and $\sim 0.68c$ and $\sim 0.002c$ in case D. As in the slower velocity cases, the turbulent velocity is subsonic, but super-Alfv\'{e}nic in most of the postshock region.

\begin{figure}[h!]
\epsscale{0.85}
\plotone{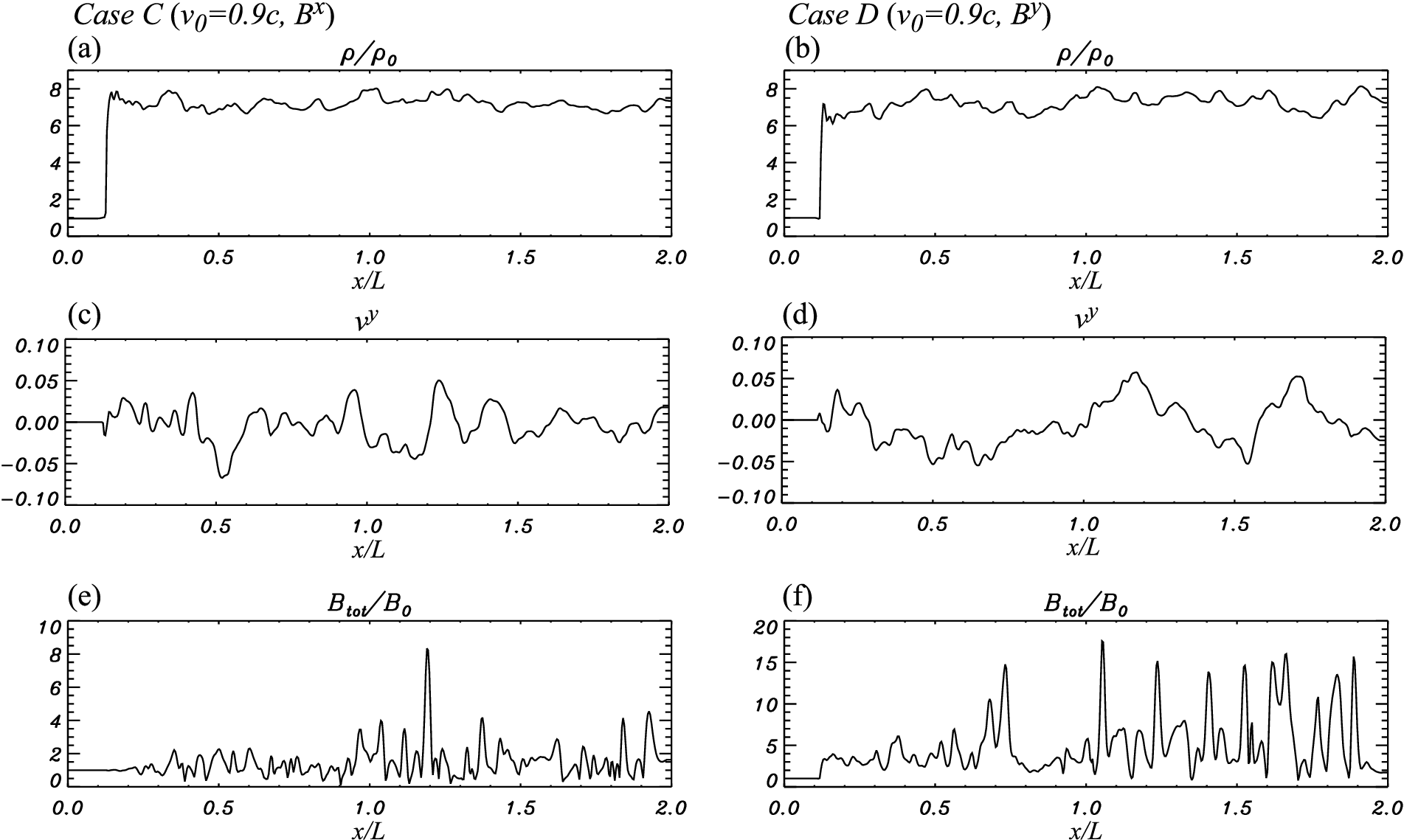}
\caption{One-dimensional profiles along the $x$-direction at $y/L=0.5$ and $t_{s}=4.4$ of ({\it panels a, b}) the density normalized by $\rho_{0}$, ({\it panels c, d}) the transverse velocity ($v^{y}$), and ({\it panels e, f}) the total magnetic-field strength ($B_{tot}$) normalized by $B_{0}$. Results for case C, involving a parallel magnetic field and $v_{0}=0.9c$, are shown on the left, whereas plots for case D, a perpendicular magnetic field and $v_{0}=0.9c$, are displayed on the right.
\label{f5}}
\end{figure}
One-dimensional profiles along the $x$-axis for cases C and D (parallel or perpendicular initial field with fast flow) are displayed in Figure 5. The shock front is located at $x=0.11L$, and the shock speed is around $v_{sh} \sim 0.43c$. This is in good agreement with analytical estimates and about 2.5 times faster than that of $v_{0}=0.4c$ cases.  The relativistic Mach number is $M_{s} \sim 26.5$ in the fast flow cases. This is more than 5 times larger than in the slow flow cases.  The density jumps by more than a factor of 7, more than the  strong shock limit in the Newtonian case (a factor of 4) because this is a mildly relativistic shock. The transverse-velocity profiles show strong fluctuations. The highest transverse velocity is about $0.07c$ in both cases. This is also a subsonic speed in the postshock region ($ <c_{s}> \simeq 0.68c$ in fast flow cases). The total magnetic field also shows substantial fluctuations. In the case of an initially perpendicular magnetic field (D), the field is compressed by nearly a factor of 3 at the shock front. The field compression behind the shock front is not identical to the density jump due to the density fluctuations. Further downstream the field is strongly amplified, reaching more than 15 times the initial field strength. In case C, involving an initially parallel magnetic field, the field is amplified purely by turbulence and locally reaches more than 8 times the initial amplitude. For a fast flow, the field amplification is stronger than in the case of a slow flow, because the stronger shock produces faster vortices in the postshock region. Because the turnover time of vortices is shorter, magnetic-field amplification is faster. Thus, the magnetic-field amplification time scale depends on the shock strength.

\subsection{Turbulent structure}
\begin{figure}[h!]
\epsscale{0.65}
\plotone{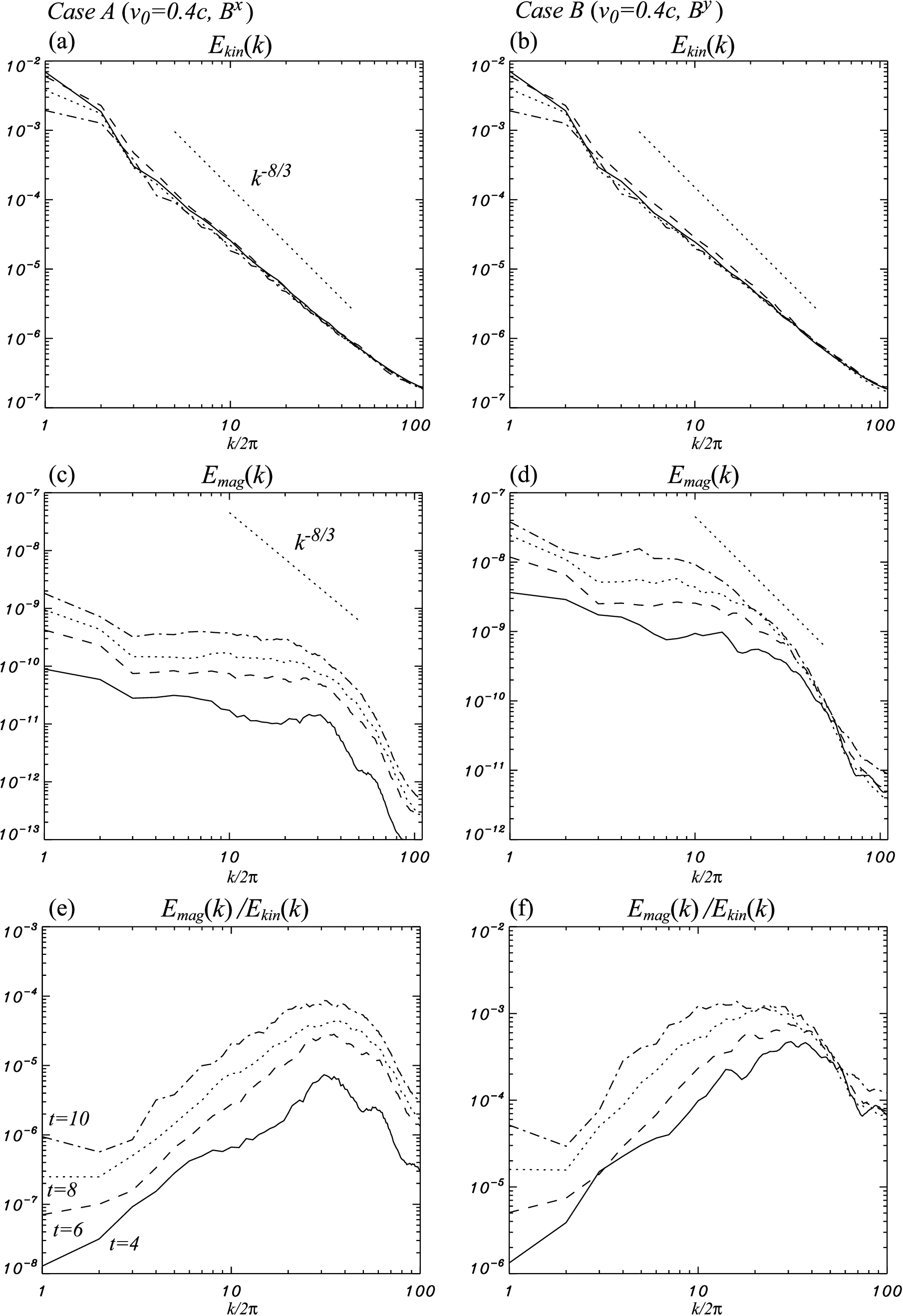}
\caption{Spherically-integrated spectra of ({\it a, b}) the kinetic energy, ({\it c, d}) the electromagnetic energy, and ({\it e, f}) the ratio of electromagnetic and kinetic energy spectra for the case A ({\it left}, parallel magnetic field, $v_{0}=0.4c$) and B ({\it right}, perpendicular magnetic field, $v_{0}=0.4c$). Different lines are for different times: $t_{s}=4$ ({\it solid}), $6$ ({\it dashed}), $8$ ({\it dotted}), and $10$ ({\it dash-dotted}). A dotted line, representing the power law $E(k) \sim k^{-8/3}$, is shown for comparison in the upper and middle panels ({\it a, b, c, d}).
\label{f6}}
\end{figure}
\begin{figure}[h!]
\epsscale{0.65}
\plotone{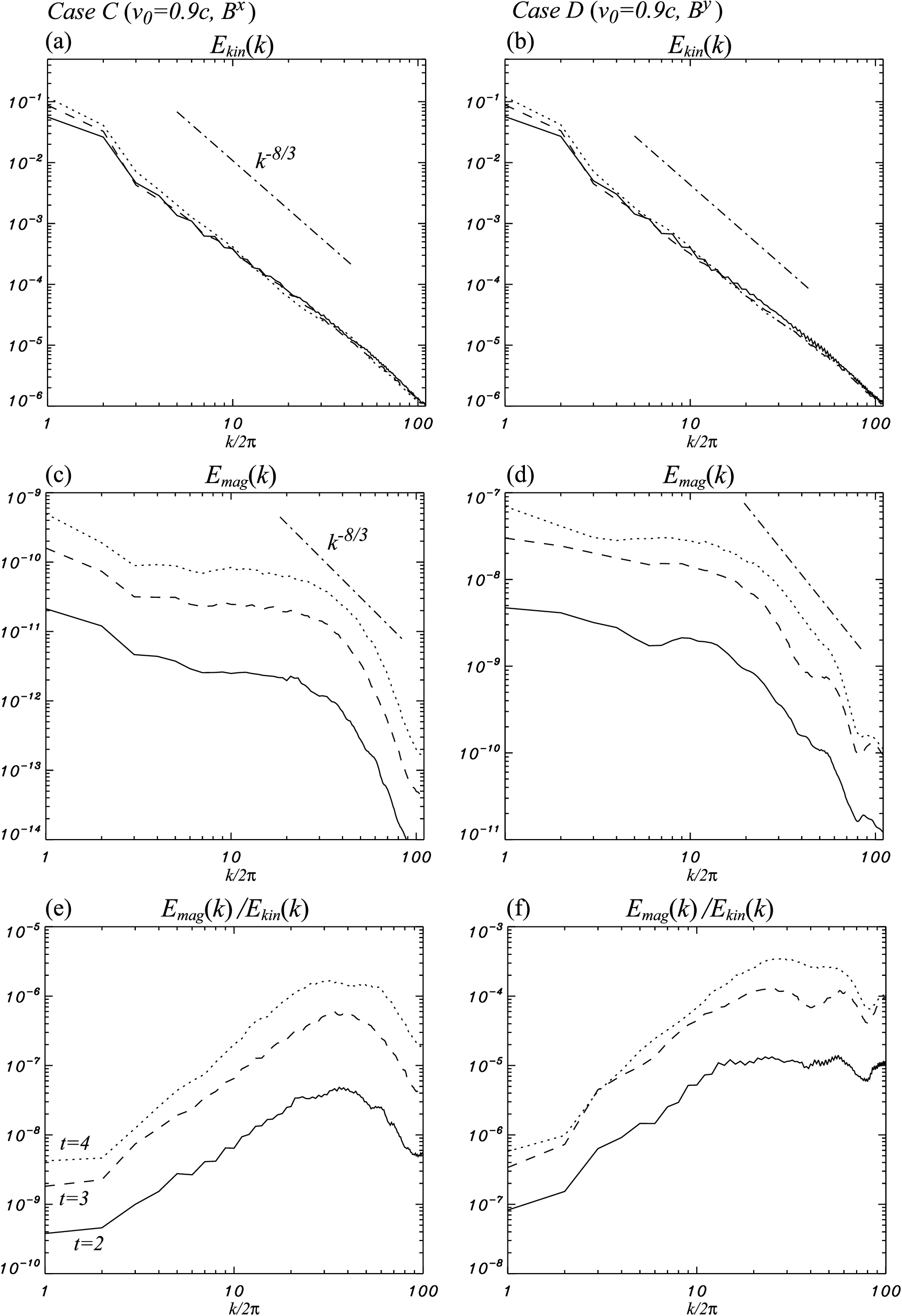}
\caption{Spherically-integrated spectra of ({\it a, b}) the kinetic energy, ({\it c, d}) the electromagnetic energy, and ({\it e, f}) the ratio of electromagnetic and kinetic energy spectra for the case C ({\it left}, parallel magnetic field, $v_{0}=0.9c$) and D ({\it right}, perpendicular magnetic field, $v_{0}=0.9c$). Different lines are for different times: $t_{s}=2$ ({\it solid}), $3$ ({\it dashed}), and $4$ ({\it dotted}). Again, a dash-dotted line representing the power law $E(k) \sim k^{-8/3}$ is shown for comparison.
\label{f7}}
\end{figure}
In order to understand the statistical properties of the turbulent fluctuations in the postshock region, it is helpful to observe their spectra. We define the kinetic energy spectrum $E_{kin}(k)$ as
\begin{equation}
\int E_{kin}(k) dk= \langle \rho \gamma (\gamma -1)\rangle ,
\end{equation}
where $\gamma$ is the Lorentz factor and the electromagnetic energy spectrum $E_{mag}(k)$ as
\begin{equation}
\int E_{mag}(k) dk= \langle T^{00}_{EM} \rangle,
\end{equation}
where $T^{00}_{EM}$ is the $00$-component of the energy-momentum tensor for the electromagnetic field (Zhang et al. 2009). Figures 6 and 7 show spherically-integrated spectra of the kinetic and electromagnetic energy for slow and fast flow parallel and perpendicular initial field cases, respectively. 

The kinetic-energy spectra almost follow a Kolmogorov spectrum in all cases, $E_{kin}(k) \propto k^{-(5/3)-(D-1)}$ with $D=2$ in two-dimensional systems. Initially the density in the preshock region is inhomogeneous with a Kolmogorov-like power spectrum, and this density spectrum still exists in the postshock region. The slope and magnitude of the kinetic-energy spectra do not change with time.

The electromagnetic-energy spectrum increases over time in all cases, implying that magnetic-field amplification is ongoing and not yet saturated. However, the overall shape of the spectra are very similar. The magnetic energy spectra are almost flat and strongly deviate from a Kolmogorov spectrum. 
Spectra flatter than a Kolmogorov spectrum are typical
of a small-scale dynamo (e.g., Childress \& Gilbert 1995; Galloway 2003; Brandenburg \& Subamanian 2005). In a small-scale dynamo, a forward cascade of magnetic energy from large scales to intermediate scales, and an inverse cascade from small scales to intermediate scales, are introduced. Therefore, the electromagnetic-field spectrum is flatter than Kolmogorov. A flat magnetic-energy spectrum is generally seen in turbulent-dynamo simulations for incompressible MHD (e.g., Schekochihin et al.\ 2004) and compressible MHD (e.g., Balsara et al.\ 2004; Balsara \& Kim 2005; Kim \& Balsara 2006). The same properties are also observed in relativistic MHD turbulence induced by the Kelvin-Helmholtz instability (Zhang et al.\ 2009).

The ratio of the electromagnetic and kinetic energy spectra is shown in the lower panels of Figures 6 and 7. This ratio increases with time in all cases. The electromagnetic energy is large at small scales but the kinetic energy remains dominant at all scales, because magnetic-field amplification by turbulence has not yet saturated. The ratio of the electromagnetic and kinetic energy spectra will probably increase up to equipartition ($\sim 1$). We would need to perform longer simulations with a larger simulation box to follow the magnetic field to saturation (from a rough estimate of magnetic energy growth, a simulation about four times longer in time and space is needed but this exceeds our present computational resources).

\subsection{Magnetic-field amplification}

Figure 8 shows the time evolution of the magnetic field, both as a volume-averaged (mean) total magnetic field and as a peak total magnetic-field strength in the postshock region (whose size increases with time as the shock propagates away from the wall), in all cases normalized by $B_0$. 
\begin{figure}[h!]
\epsscale{0.65}
\plotone{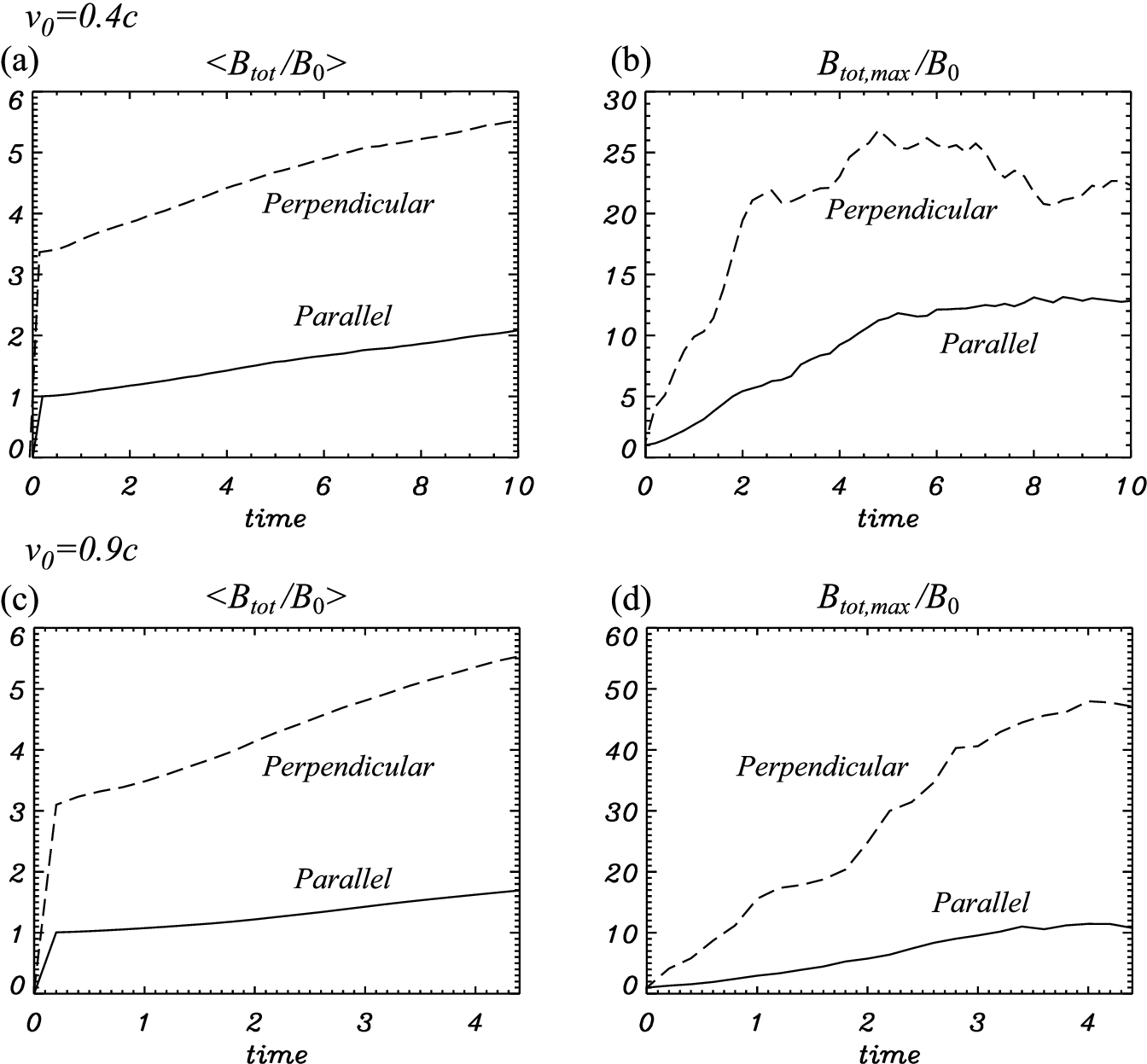}
\caption{Time evolution of ({\it a, c}) the volume-averaged total magnetic field and ({\it b, d}) the maximum total magnetic-field strength in the postshock region normalized by $B_0$ for the case of lower flow velocity ($v_{0}=0.4c$: {\it upper panels}) and higher velocity ($v_{0}=0.9c$: {\it lower panels}). Different lines are for different initial magnetic field orientation: parallel ($B^x$: {\it solid}) and perpendicular ($B^y$: {\it dashed}) with respect to the shock propagation direction. \label{f8}}
\end{figure}
Note that the mean magnetic-field strength is still increasing when the simulation was terminated; apparently the magnetic-field strength takes some time to saturate. The mean postshock magnetic field is stronger for the perpendicular shock case ($B^{y}$), for which we observe an increase by more than a factor of 5 compared to $B_0$. For the parallel shock case ($B^{x}$) an amplification by about a factor of 2 is observed. In the perpendicular shock case, the perpendicular magnetic field is compressed by a factor of 3 at the shock, and additional magnetic field amplification by turbulent motion in the postshock region is almost the same as for the parallel magnetic field case (see in Fig. 9). In both cases, the peak field strength is much larger than the mean magnetic field, about $13$ and $26$ times for the slow flow velocity and $12$ and $49$ times for the fast velocity, where the two numbers are for the parallel and perpendicular cases, respectively. The time scale for magnetic-field amplification in the fast-flow case is about half that in the slow-flow case, because vortices with faster velocity can be produced.

\begin{figure}[h!]
\epsscale{0.65}
\plotone{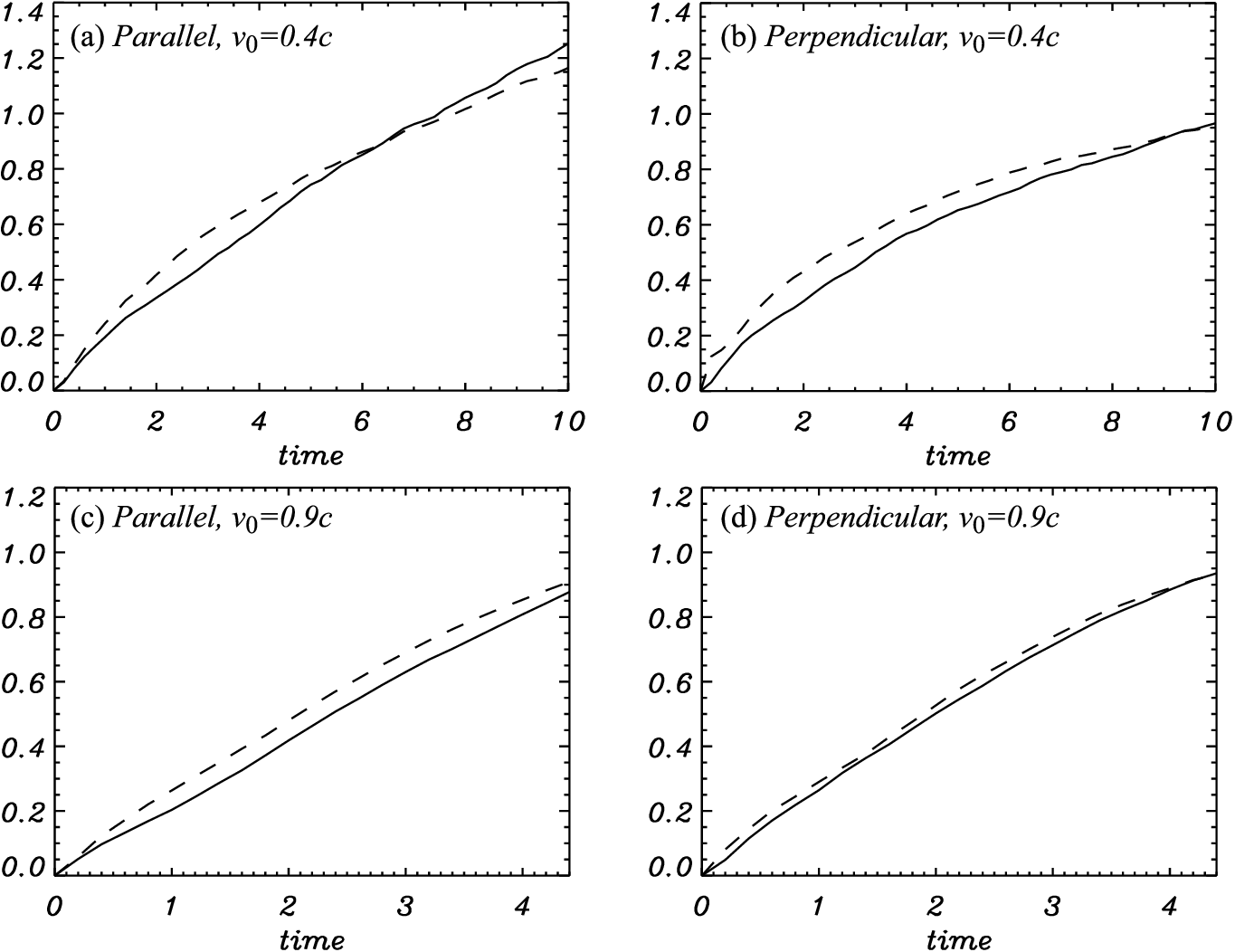}
\caption{Time evolution of the volume-averaged root mean square (rms) fluctuation amplitudes of $\delta B^{x}$ ({\it solid lines}) and $\delta B^{y}$ ({\it dashed lines}) in the postshock region normalized by the mean magnetic field strength in postshock region for the cases A, B, C, and D. \label{f9}}
\end{figure}
To investigate the amount of magnetic-field amplification via turbulent motion, the time evolution of the mean fluctuation amplitudes of $\delta B^{x}$ and $\delta B^{y}$ in the postshock region normalized by the mean magnetic-field strength ($B_{s} \simeq 4.5 \times 10^{-3}$ for the parallel-field case, $\simeq 0.014$ for the perpendicular-field case) is shown in Figure 9. In all cases, the fluctuation amplitude increases with time, showing no sign of saturation. Each fluctuating magnetic field component grows similarly, implying that the turbulence amplifies each magnetic field component equally. The growth rate is independent of the original magnetic-field direction. Therefore, the amplification of magnetic field by turbulent motion does not depend on the direction of a homogeneous magnetic field. We should note that this magnetic field fluctuation amplitude is normalized by the mean magnetic field in the postshock region, which is about 3 times larger in the perpendicular-field case than in the parallel-field case on account of magnetic field compression by the shock. Therefore, the total magnetic-field amplification is larger for a perpendicular homogeneous field.

\section{Summary and Discussion}

We have performed two-dimensional relativistic MHD simulations of the propagation of a mildly relativistic shock through an inhomogeneous medium. We have shown that the postshock region becomes turbulent owing to the preshock density inhomogeneity, and the magnetic field is strongly amplified by the turbulent motion in the postshock region. The amplified magnetic field evolves into filamentary structures in these two-dimensional simulations. This is consistent with earlier non-relativistic studies (e.g., Balsara et al.\ 2001, 2004; Giacalone \& Jokipii 2007; Inoue et al.\ 2009). The magnetic energy spectrum is flatter than Kolmogorov, which is typical for a small-scale dynamo. We also found that the total magnetic-field amplification from the preshock value depends on the direction of the homogeneous magnetic field, and the time scale of magnetic field growth depends on the shock strength. The mean magnetic field strength in the postshock region is still increasing when the simulations were terminated. Therefore, longer simulations with a longer simulation box are needed to follow the magnetic field amplification to saturation.

In this paper we have performed simulations in two-dimensional geometry as a first step toward understanding magnetic-field amplification by turbulence in the relativistic MHD regime. However, in general the turbulence structure, for example the slope of the Kolmogorov-like power spectrum, is different in two and three dimensions. To more realistically analyze three-dimensional phenomena, we will extend the current investigation to three-dimensional simulations in future work. 

Recently, the Fermi satellite observed GRB 080916C which showed a featureless smoothly-joint broken power-law spectra covering 6-7 decades in energy (Abdo et al. 2009). Zhang \& Pe'er (2009) analyzed this burst in detail and argued that the non-detection of a thermal component strongly suggests that the outflow is Poynting-flux dominated with magnetization $\sigma \gtrsim 15$ (the Poynting-to-kinetic energy ratio) at the central engine. Therefore, at least some GRB outflows appear to be magnetically dominated. In our simulations, we assume the preshock medium is weakly magnetized. If we consider internal shocks in GRBs, which are internal collisions within relativistically propagating shells, the magnetization in the preshock medium can cover a wide range of values. In particular, it is expected that in the strongly-magnetized (high $\sigma$) regime the MHD 
turbulence is anisotropic and has a different scaling in the directions along and perpendicular to the homogeneous field, $E(k_{\perp}) \propto k^{-5/3}_{\perp}$ (Kolmogorov-type) and $E(k_{\parallel}) \propto k^{-2}_{\parallel}$ (Goldreich \& Sridhar 1995; Cho \& Lazarian 2003). While the kinetic power in the turbulence quickly decreases with $k$, the power in the magnetic field does not significantly decrease with $k$. As a result, eddies of smaller scales are more stretched and appear elongated along the magnetic field. Current simulations do not show any anisotropy because the magnetic field in the turbulent postshock medium is still weak. MHD turbulence in the highly-magnetized (high $\sigma$) regime is not well studied. A comprehensive study of relativistic turbulence with a wide range of magnetization is now underway and will be presented in future publications.

In the internal shock model, the GRB central engine ejects an unsteady outflow that can be visualized as discrete magnetized shells with variable Lorentz factor and mass loading. These magnetized shells collide with each other at the conventional internal shock radius, 
$R_{IS} \simeq 2 \gamma^{2} c \delta t \simeq 10^{15}  \gamma^{2}_{2.5} \delta t$ cm, 
where $\gamma_{2.5}=\gamma/10^{2.5}$. Hereafter the convention $Q_{s} = Q/10^{s}$. The gamma-ray emission radius $R$ can be in principle in the range of $10^{11}$-$10^{17}$ cm (between the photosphere radius and the external shock deceleration radius). Various observational constraints suggest that the non-thermal emission radius is at $R \geq 10^{14}$ cm (e.g., Kumar et al. 2007; Racusin et al. 2008; Shen \& Zhang 2009; Zhang \& Pe'er 2009). The thickness of the discrete shell is  
$\Delta' \simeq R / \gamma \simeq 10^{12}  R_{15} \gamma_{2.5}^{-1}
$ cm in the comoving frame.
From a turbulence model of GRBs (Narayan \& Kumar 2009; Lazar et al. 2009), the size of an eddy in the comoving frame is 
$ R_{e} \simeq  f^{1/a} R / \gamma \gamma_{t}$, where $f$ is a filling factor and $\gamma_{t}$ is a random Lorentz factor for an energy-bearing eddy. Lazar et al. (2009) suggest that $a$ is between 2 and 3. Narayan \& Kumar (2009) have pointed out that $R_{e} = R/\gamma \gamma_{t}$ if one requires that the eddies be of the maximal causally allowed size. Our simulation results show that eddies have different sizes and randomly oriented flow directions. Clearly $R_{e}$ will be very small.

There is a question as to whether a fluid treatment of the GRB plasma is justified. The same question arises in supernova remnants, to which the standard answer is that magnetization of the plasma ensures a short effective mean free path.
From the detailed discussion in Zhang \& Yan (2010), the strong collision radius for Coulomb collisions may be defined by $e^{2}/r_{col} \sim kT$ so that $r_{col} \sim e^{2} /kT \sim 10^{-3}/T$. The comoving collision mean free path of electrons can be estimated as
$l'_{e,col} = (n'_{e} \pi r^{2}_{col})^{-1} \simeq 10^{17} $ cm, 
where $n'_{e}$ is the electron number density in the comoving frame.
In order to have the plasma in the collisional regime, one needs to require $l'_{e,col} < \Delta'$. Therefore GRB shocks must be collisionless. However, the GRB fluid stays together through magnetic interactions. A more relevant mean free path would be a magnetic gyro-radius and plasma skin depth. The gyro(cyclotron)-radii in the comoving frame are
$r'_{B,e} = \gamma_{e} m_{e} c^{2} / e B' \sim 1 $ cm
for electrons (where $\gamma_{e}$ is the electron Lorentz factor and $B'$ is the comoving magnetic field strength), and
$r'_{B,p} = \gamma_{p} m_{p} c^{2} / e B' \sim 10^{3} $ cm 
for protons (where $\gamma_{p}$ is the proton Lorentz factor). 
The comoving relativistic plasma skin depths are 
$ \delta'_{e} = c / \omega'_{p,e} =  \bar{\gamma_{e}} m_{e} c^{2} / 4 \pi n'_{e} e^{2} \sim 10^{2}$ cm for electrons, and 
$ \delta'_{p} = c / \omega'_{p,p} = \bar{\gamma_{p}} m_{p} c^{2} / 4 \pi n'_{p} e^{2}  \sim 10^{3} $ cm
for protons. Here $\bar{\gamma_{e}}$ and $\bar{\gamma_{p}}$ denote the mean Lorentz factor of the relativistic electrons and protons, respectively. Physically, plasma skin depths are relevant to  non-magnetized ejecta. For magnetized ejecta, the plasma skin depths are relevant in the direction parallel to the magnetic field lines. The gyro-radii are more relevant in the direction perpendicular to the magnetic field lines. Both magnetic gyro-radii and plasma skin depths are $\ll \Delta'$. Therefore this justifies a fluid treatment of the GRB plasma.

Our simulations show filamentary magnetic-field structure. A turbulent plasma with fast-moving magnetic filaments is likely a site for second order Fermi acceleration of charged particles and a source of high energy cosmic rays. The spatial variability and large fluctuations of the turbulent magnetic field also have implications for nonthermal radiation with possibly observable consequences. A simple light curve (the variability amplitude) can be obtained from integrating $B_{\perp}^{2}$ with respect to the light of sight measured in the laboratory frame over the postshock region as a proxy of the electron emissivity. 
Within the GRB context, turbulence-powered lightcurves were calculated by Narayan \& Kumar (2009) and Lazar et al. (2009). In their calculations, the relativistic turbulent eddies have a characteristic Lorentz factor, and the electron emissivity can be characterized by $\gamma_{e}^{2} B'^{2}$ along with a Lorentz boost  due to the turbulent motion. As a result, the lightcurve shows a peak when an eddy enters into the line of sight (similar to synchrotron radiation). This can produce significant variability. In our case, the turbulence is mildly relativistic, so that at any epoch, one sees the contribution from many eddies since the Lorentz boost eddy effect is not significant. For our current simulations we would expect the eddy emission to average out, and as a result we would expect a roughly flat lightcurve without significant variability. 
A self-consistent calculation of nonthermal radiation, which includes the acceleration of nonthermal electrons and the amplification of magnetic fields would be highly valuable.

\acknowledgments

Y.M. thanks UNLV for hospitality during the preparation of part of the work. Y.M., P.H., and K.N. acknowledge  partial support by NSF awards AST-0506719, AST-0506666, AST-0908010, and AST-0908040, and NASA awards NNG05GK73G, NNX07AJ88G, and NNX08AG83G. J.N. is partially supported by MNiSW research project N N203 393034, and The Foundation for Polish Science through the HOMING program, which is supported by a grant from Iceland, Liechtenstein, and Norway through the EEA Financial Mechanism.
BZ acknowledges NSF award AST-0908362 and NASA awards NNX09AT66G and NNX10AD48G for support. The simulations were performed on the Columbia Supercomputer at the NAS Division of the NASA Ames Research Center, the SGI Altix (cobalt) at the National Center for Supercomputing Applications in the TeraGrid project supported by the NSF, and the Altix3700 BX2 at YITP in Kyoto University.

\appendix

\section{Calculation of shock wave propagation speed}

In our simulations, simulation box is the contact discontinuity frame and plasma moves into the contact discontinuity (the boundary at $x=x_{max}$) from the left side with a flow speed $v_{x}=v_{0}=0.4c$ or $0.9c$ (a Lorentz factor of $\gamma_{0}=1.1$ or $2.3$). The shock wave moves away from contact dicontinuity in the $-x$ direction. From the relativistic shock jump conditions (Blandford \& McKee 1976) in the ``flow'' (primed) frame, i.e., the frame of the plasma injected at the boundary of $x=x_{min}$ (preshock frame), the shock Lorentz factor is given by
\begin{equation}
\gamma'^{2}_{sh} = { (\gamma_{0}+1) [\Gamma (\gamma_{0}-1)+1]^{2} \over
\Gamma (2-\Gamma) (\gamma_{0}-1) +2}.
\end{equation}
The shock propagation speed in the contact discontinuity frame is obtained from the Lorentz transformation of the velocity as
\begin{equation}
v_{sh}= {v'_{sh} - v_{0} \over 1- v'_{sh} v_{0}},
\end{equation}
where $v'_{sh}$ is the shock propagation speed in the flow frame.
Using an adiabatic index $\Gamma=5/3$, the shock propagation speed in the contact discontinuity frame is $v_{sh}=0.17c$ for the $v_{0}=0.4c$ case and $v_{sh}=0.43c$ for the $v_{0}=0.9c$ case, respectively.

\end{document}